\documentclass[
 aip,
 amsmath,amssymb,
reprint,
]{revtex4-1}

\usepackage{graphicx}
\usepackage{dcolumn}
\usepackage{bm}

\usepackage[utf8]{inputenc}
\usepackage[T1]{fontenc}
\usepackage{mathptmx}
\usepackage{etoolbox}
\usepackage{adjustbox}
\usepackage{wrapfig}
\usepackage{subcaption}

\makeatletter
\def\@email#1#2{%
 \endgroup
 \patchcmd{\titleblock@produce}
  {\frontmatter@RRAPformat}
  {\frontmatter@RRAPformat{\produce@RRAP{*#1\href{mailto:#2}{#2}}}\frontmatter@RRAPformat}
  {}{}
}%
\makeatother
\begin{document}

\preprint{AIP/123-QED}

\title[Manuscript Valerio Digiorgio APL]{$\mbox{Rb}_2\mbox{Ti}_2\mbox{O}_5$ : a layered ionic conductor at the sub-micrometer scale}
\author{Valerio Digiorgio}
 \altaffiliation[Presently at ]{ETH Zürich, Institute for Quantum Electronics, Department of Physics, Zürich, Switzerland}
 
 \author{Karen Sobnath}%

\affiliation{ 
Université Paris Cité, Laboratoire Matériaux et Phénomènes Quantiques, CNRS, UMR 7162, 75013 Paris, France
}%
\author{Maria Luisa Della Rocca}%

\affiliation{ 
Université Paris Cité, Laboratoire Matériaux et Phénomènes Quantiques, CNRS, UMR 7162, 75013 Paris, France
}%

\author{Clément Barraud*}%
  \email{clement.barraud@u-paris.fr}
\affiliation{ 
Université Paris Cité, Laboratoire Matériaux et Phénomènes Quantiques, CNRS, UMR 7162, 75013 Paris, France
}%

\author{Rémi Federicci}
\affiliation{%
 Pioniq Technologies, 6 rue Jean Calvin, 75005 Paris, France 
}%

\author{Armel Descamps-Mandine}
\altaffiliation[Presently at ]{CNRS, Centre de Microcaractérisation Raimond Castaing, 3 Rue Caroline Aigle, 31400 Toulouse, France}
\affiliation{%
 LPEM-ESPCI Paris, PSL Research University, CNRS, Sorbonne Université UPMC,
10 rue Vauquelin, 75005 Paris, France
}%
 
\author{Brigitte Leridon*}
  \email{brigitte.leridon@pioniq-technologies.com}
\affiliation{%
 Pioniq Technologies, 6 rue Jean Calvin, 75005 Paris, France 
}%
\affiliation{%
 LPEM-ESPCI Paris, PSL Research University, CNRS, Sorbonne Université UPMC,
10 rue Vauquelin, 75005 Paris, France
}%

\date{\today}

\begin{abstract}
Over the past few years, ionic conductors have gained a lot of attention given the possibility to implement them in various applications such as supercapacitors, batteries or fuel cells as well as for resistive memories. Especially, layered two-dimensional (2D) crystals such as h-BN, graphene oxide and $\mbox{MoSe}_2$ have shown to provide unique properties originating from the specific 2D confinement of moving ions. Two important parameters are the ion conductivity and the chemical stability over a wide range of operating conditions.
In this vein, $\mbox{Rb}_2\mbox{Ti}_2\mbox{O}_5$ has been recently found displaying remarkable properties such as superionic conduction and colossal equivalent dielectric constant.
Here, a first approach to the study of the electrical properties of layered $\mbox{Rb}_2\mbox{Ti}_2\mbox{O}_5$ at the 100-nanometer scale is presented. Characterizations by means of micro-Raman spectroscopy and atomic force microscope (AFM) measurements of mechanically exfoliated RTO nanocrystals via the so-called adhesive-tape technique are reported. Finally, the results of electrical measurements performed on an exfoliated RTO nanocrystals are presented, and are found to be consistent with the results obtained on macroscopic crystals.

\end{abstract}

\maketitle

%

Ionic solid-state conductors present the capability to conduct ions through their -organic or inorganic- structure \cite{Rickert1978}.  Remarkably, in some 2D lamellar crystals such as, for instance, graphite or h-BN,  ions can easily move in between the stacked atomic planes\cite{Chen2022,Yu2023,Park2021}. These migration planes that act as 2D-conduction channels can reveal unique features such as either anisotropic or isotropic high ionic conductivity\cite{Chen2022,Yu2023,Park2021} and metal-insulator transition of the host material\cite{Han2023}. Technologically, those layered materials are also promising with respect to the great diversity of compatible fabrication processes and deposition techniques that are available\cite{Balaish2021}. 
To date, a variety of solid-state inorganic materials have been investigated \cite{Famprikis_2019_NatureMat}, among them perovskite-type materials. The perovskite-derived lamellar material $\mbox{Rb}_2\mbox{Ti}_2\mbox{O}_5$ (in the following RTO), belonging to the $\mbox{M}_2\mbox{Ti}_2\mbox{O}_5$ (with M = Li, Na, K, Rb, Cs, Fr) family, whose structure was identified in the 1960’s \cite{Andersson_five_1960,andersson_The_1961}, has recently been found to display remarkable properties, which make it interesting for different applications in the energy field. Its colossal equivalent dielectric constant (up to $10^9$ at room temperature), high ionic conductivity (up to $1\,\mbox{mS.cm}^{-1}$), and very low electronic conductivity ( $\sigma_{elec.}\leq 10^{-8} \mbox{S.cm}^{-1}$) are particularly interesting for its use as a solid electrolyte for the realization of devices for energy storage \cite{PhysRevMaterials.1.032001}.
In addition, memory effects were shown to be present in RTO single crystals, which exhibit a frequency-dependent I-V curve when biased upon large voltages ($V_{bias}>2\,V$), thus also making RTO a memristive\cite{chua_resistance_2011} material \cite{doi:10.1063/1.5036841}.

In the present letter, we investigate RTO electronic and structural properties at the low-dimensional limit. By taking advantage of the material's lamellar structure, we demonstrate the feasibility of RTO mechanical exfoliation and transfer onto substrates. We probe transport properties of micrometer size flakes of exfoliated RTO of few hundreds nanometers thickness, showing that memory effects observed in single crystals are preserved at the low dimensional scale. Our findings make low-dimensional RTO a high potential material for nanoscale memory development.

\begin{figure*}
 \includegraphics[scale=0.7]{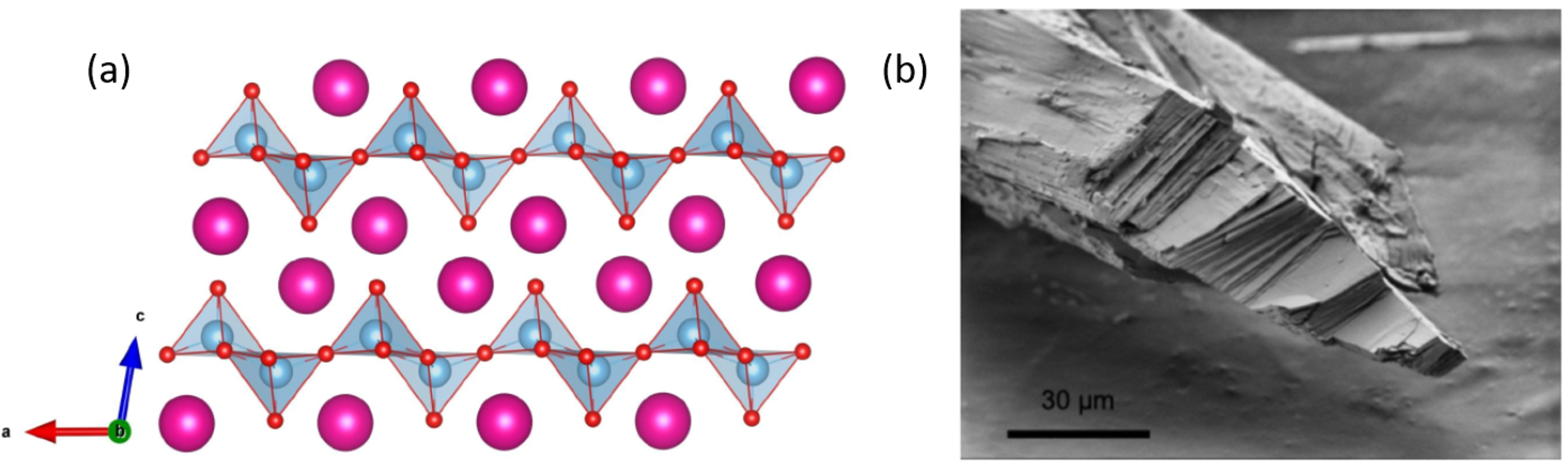}
 \caption{\label{MEB}(a) Cristallographic structure of the RTO crystal as determined previously in Ref.\cite{PhysRevMaterials.1.032001}. Pink atoms are Rb atons. Blue atoms are Ti atoms. Red atoms are O atoms. ab-planes corresponds to the atomic planes. The c-axis represents the out-of-plane direction. (b) Scanning electron microscope image of the edge of a RTO bulk crystal evidencing its lamellar structure along the c-axis.}%
\end{figure*}

Little information has been reported to date in the scientific literature regarding RTO and its properties. Two early reports exist on the synthesis and crystal structure of RTO, which was stated to be consistent with the C2/m space group \cite{Andersson_five_1960,andersson_The_1961}. The unit cell of RTO consists in a two-dimensional structure made of alternate layers of Ti-O (namely $(\mbox{Ti}_2\mbox{O}_5)^{2-}$ ) and Rb (namely $\mbox{Rb}^+$) atoms stacked along the c axis. The $(\mbox{Ti}_2\mbox{O}_5)^{2-}$ planes are composed of staggered upside and downside oxygen pyramids with Ti atoms at the center, that are therefore fivefold coordinated A schematic of the structure is displayed in Fig.\ref{MEB}a. The lamellar structure of the crystal, which was confirmed by HRTEM and SEM images \cite{Federicci_The_2017}, makes RTO particularly suitable for mechanical exfoliation\cite{doi:10.1073/pnas.0502848102} as revealed in Fig.\ref{MEB}b.

A recent work on RTO has demonstrated superionic conduction properties stemming from the electrolytic nature of the material allowing ionic transport. The ionic transport mechanism is not fully understood, but, as far as it is understood, water absorption is a crucial factor\cite{PhysRevMaterials.1.032001,DESOUSACOUTINHO2021115630}. It is already established that lamellar oxides such as transition metal oxyde are prone to water intercalation \cite{doi:10.1063/1.1610805}. RTO crystals have been shown to spontaneously hydrate when exposed to ambient atmosphere. Moreover, part of the intercalated water molecules experience dissociation producing protons and hydroxide ions, which triggers the "activation" of the material as a superionic conductor\cite{DESOUSACOUTINHO2021115630}. 

The equivalent dielectric constant and the electrical polarization of spontaneously hydrated RTO were shown to reach unprecedented values such as $10^9$ and $0.1 \mbox{C.cm}^{-2}$, respectively, which was attributed in part to the high ionic conduction of the order of $10^{-3} S.cm^{-1}$ at 273 K (and $10^{-4} S.cm^{-1}$ at 300 K), together with extremely low electronic conductivity (lower than $10^{-8} S.cm^{-1}$) at low voltage, but also to the phase change inside the material under voltage application dubbed virtual cathode \cite{DESOUSACOUTINHO201972}. These combined properties make the material simultaneously a superionic conductor and an extremely good electronic insulator \cite{PhysRevMaterials.1.032001}.

In addition RTO single crystals were found to display a memristive behaviour, meaning that their impedance at a given time is a function of the charge that has flown through them.
As a consequence typical I-V characteristics of RTO single crystals exhibit a frequency-dependent behavior\cite{doi:10.1063/1.5036841}. For low frequencies (around 1\,mHz) the electrical measurements show non-linear and non-transverse - the two branches do not cross at the origin, hysteretic loops that are pinched in zero. Increasing the frequency of the applied voltage at about 500\,mHz, the I-V characteristics hysteresis shrinks and then the characteristics adopts a ohmic-like behavior with a single-valued conductivity of about $10^{-4}\,\mbox{S.cm}^{-1}$.  


Bulk crystals of RTO were grown using the technique described in Ref. [\onlinecite{Federicci_The_2017}]. TiO$_2$ (rutile) and RbNO$_3$ powders were placed in a Pt crucible in a 1:1 molar ratio and heated up to 1100\,°C in a furnace. The product was then slowly cooled down and transparent needles corresponding to crystals of RTO, typically 1-2 millimeters-long and 100-micrometers-wide were obtained. The single crystals were subsequently isolated and analyzed by X-ray diffraction confirming the C2/m group structure as established in [\onlinecite{Federicci_The_2017}]. A typical scanning electron microscope image of a single crystal is shown in Fig. \ref{MEB}b, where the lamellar structure of the solid compound is clearly visible.

The adhesive-tape exfoliation technique was adopted to perform micro-mechanical cleavage of RTO crystals in order to obtain very thin nanocrystals of typical lateral dimensions of tens of micrometers and of typical thicknesses hundreds of nanometers. The exfoliated nanocrystals were subsequently characterized by atomic force microscopy (AFM) and micro-Raman spectroscopy.

AFM measurements were conducted at ambient conditions on a WITEC alpha300R microscope to establish the typical size and height and the surface aspects of the RTO nanocrystals. Overall, 74 nanocrystals were investigated and exhibited either terrace-like structures or homogeneous thicknesses with surface roughnesses below  1 nm. The analyzed nanocrystals had thicknesses ranging from 50\,\mbox{nm} to 1000\,\mbox{nm} with an average value of about  300\,\mbox{nm}.

\begin{figure*}
 \includegraphics[scale=0.8]{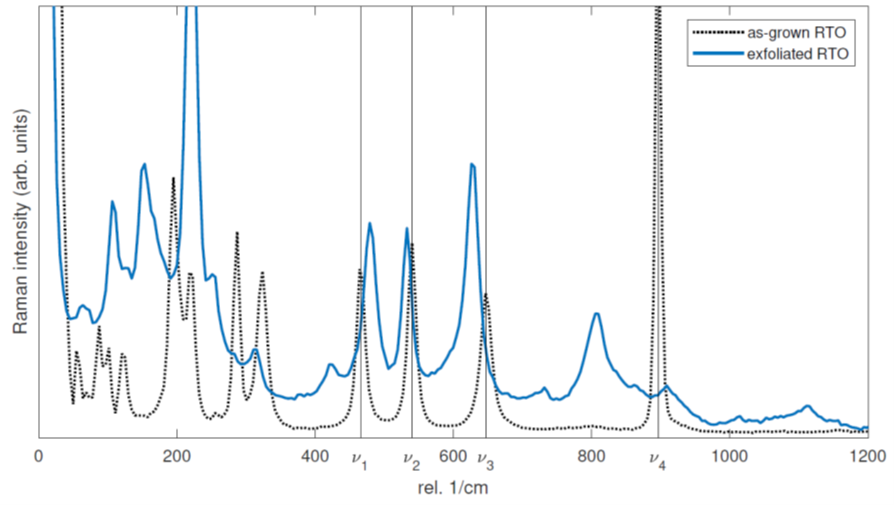}%
 \caption{\label{Ram1}Blue solid line: Raman spectrum of a single RTO exfoliated nanocrystal deposited on a Si/SiO2 (300 \,nm) substrate (flake \#0). Black dotted line: Reference Raman spectrum of a bulk RTO crystal. Four computed phonon modes\cite{Federicci_The_2017} are indicated by the vertical black lines and referred as $\nu1$, $\nu2$, $\nu3$ and $\nu4$. The peak displacements observed in the spectrum of exfoliated samples is consistent with previous observations in hydrated bulk samples.  }%
\end{figure*}

\begin{figure*}
 \includegraphics[scale=0.8]{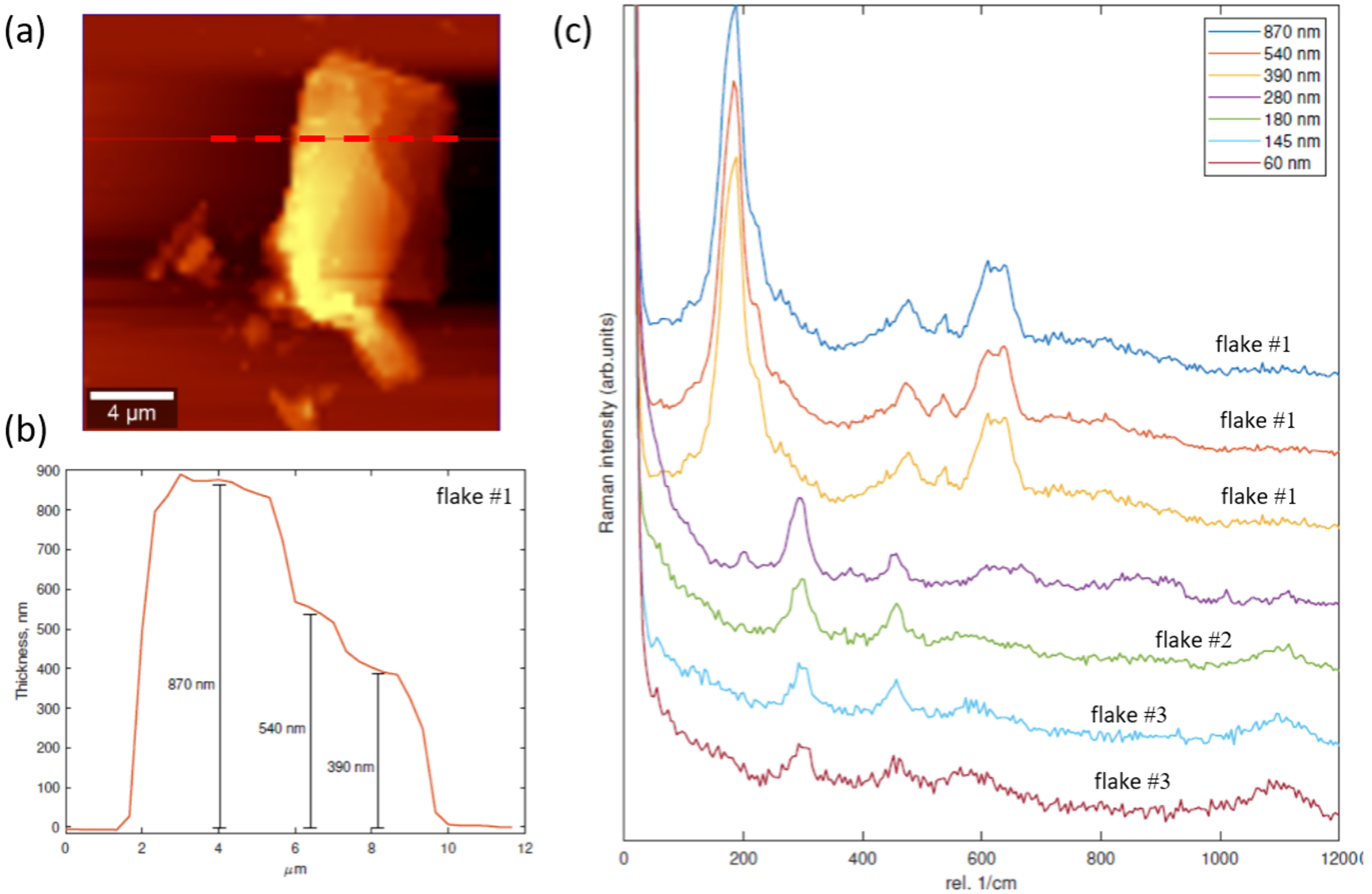}%
 \caption{\label{Ram2} (a) AFM image of an exfoliated RTO flake presenting some terraces of different thicknesses. (b) AFM profile extracted along the red dashed line displayed in (a). (c) Three top curves: Raman spectra corresponding to various thicknesses of exfoliated flake \#1 from (a), presenting step-like structures. Three bottom curves : Raman spectra corresponding to various thicknesses of exfoliated flake \#2 having step-like structure.  Intermediate purple curve: Raman spectrum for exfoliated uniform sample \#3. The variation among the spectra appears not to be thickness-dependent but rather sample-dependent. The different curves are vertically shifted for clarity}%
\end{figure*}

Exfoliated RTO nanocrystals were also studied through micro-Raman spectroscopy to investigate possible local structural changes from known bulk RTO crystals. Stokes Raman spectra were collected both from as-grown RTO bulk crystals and from the exfoliated nanocrystals with excitation wavelength of 532\,nm and laser power ranging between 5\,mW and 10\,mW. The results of the measurements are compared in Fig.\,\ref{Ram1}. The four phonon modes indicated as $\nu1$, $\nu2$, $\nu3$ and $\nu4$, observed for bulk RTO crystals appear at 466\,$\mbox{cm}^{-1}$, 540\,$\mbox{cm}^{-1}$, 647\,$\mbox{cm}^{-1}$ and 897\,$\mbox{cm}^{-1}$ respectively. These values are in agreement with values both simulated and measured in Ref. [\onlinecite{Federicci_The_2017}]. By contrast, on the RTO nanocrystal spectrum one can notice the suppression of the $\nu4$ peak and the shift of the three remaining peaks. Also, additional peaks appear in the region between $\nu3$ and $\nu4$. The same features were previously reported for bulk RTO crystals in Ref. [\onlinecite{Federicci_The_2017}] and were attributed to an activation process induced by a thermal annealing followed by a rehydration of the RTO crystal\cite{DESOUSACOUTINHO2021115630}. In the present study, RTO nanocrystals also appeared as "activated" from their Raman signatures. This observation is consistent with previous reports since they were hydrated by exposition to the atmosphere for several hours\cite{DESOUSACOUTINHO2021115630} after being exfoliated from bulk crystals. 

AFM and Raman characterizations measurements were then conducted in combination in order to investigate possible size-dependent modification in the Raman response of RTO nanocrystals. No evidence of such dependence was observed for a range of thicknesses from 60\,nm to 870\,nm. Raman spectra collected from distinct samples were slightly different but no correlation with the nanocrystals size or thickness was identified. As a general rule, two different samples with comparable thickness could be found to present different Raman signatures sharing common features notably due to different level of hydration as explained before \cite{DESOUSACOUTINHO2021115630}. At the same time, Raman spectra extracted from the same nanocrystal were found to share exactly the same features regardless of the RTO thickness in the studied range. As an example of this, several spectra were collected from three different flakes.  Two of them had a step-like structure and were suitable for analysing Raman spectra collected from regions with different thicknesses. From flake \#1, spectra were collected from terraces corresponding to  60\,nm, 145\,nm, and 180\,nm. From sample \#3, three spectra were extracted, corresponding to thicknesses of 390\,nm, 540\,nm, and 870\,nm. One additional spectrum was collected from sample \#2 that was 280\,nm-thick. These spectra are presented in Fig. \ref{Ram2}c. The signal intensity decreases with the thickness of the flakes. The emergence of a peak around 1100\,$\mbox{cm}^{-1}$ occurs in the thinnest sample, but it corresponds to the increasing signal coming from the Si substrate. Fig. \ref{Ram2}c therefore confirms that Raman spectra do not show a particular thickness-dependence, but generally, a flake-dependence since they share common characteristics if collected from the same nanocrystal for the studied thickness range.

All these findings point toward a clear preservation of the RTO structure when exfoliated because the bulk spectra are recovered, but changes in the $\nu_1-\nu_4$ peak structure of the Raman spectra from one sample to another are observed in the exfoliated samples as well as in the bulk. The role of water absorption as a possible cause is extremely likely as reported before in Ref.\onlinecite{DESOUSACOUTINHO2021115630}. In particular, a different time of exposure to water vapor and uncontrolled atmosphere will lead to a different local crystal structure and hence to a different Raman spectrum. 

\begin{figure*}
 \includegraphics[scale=0.6]{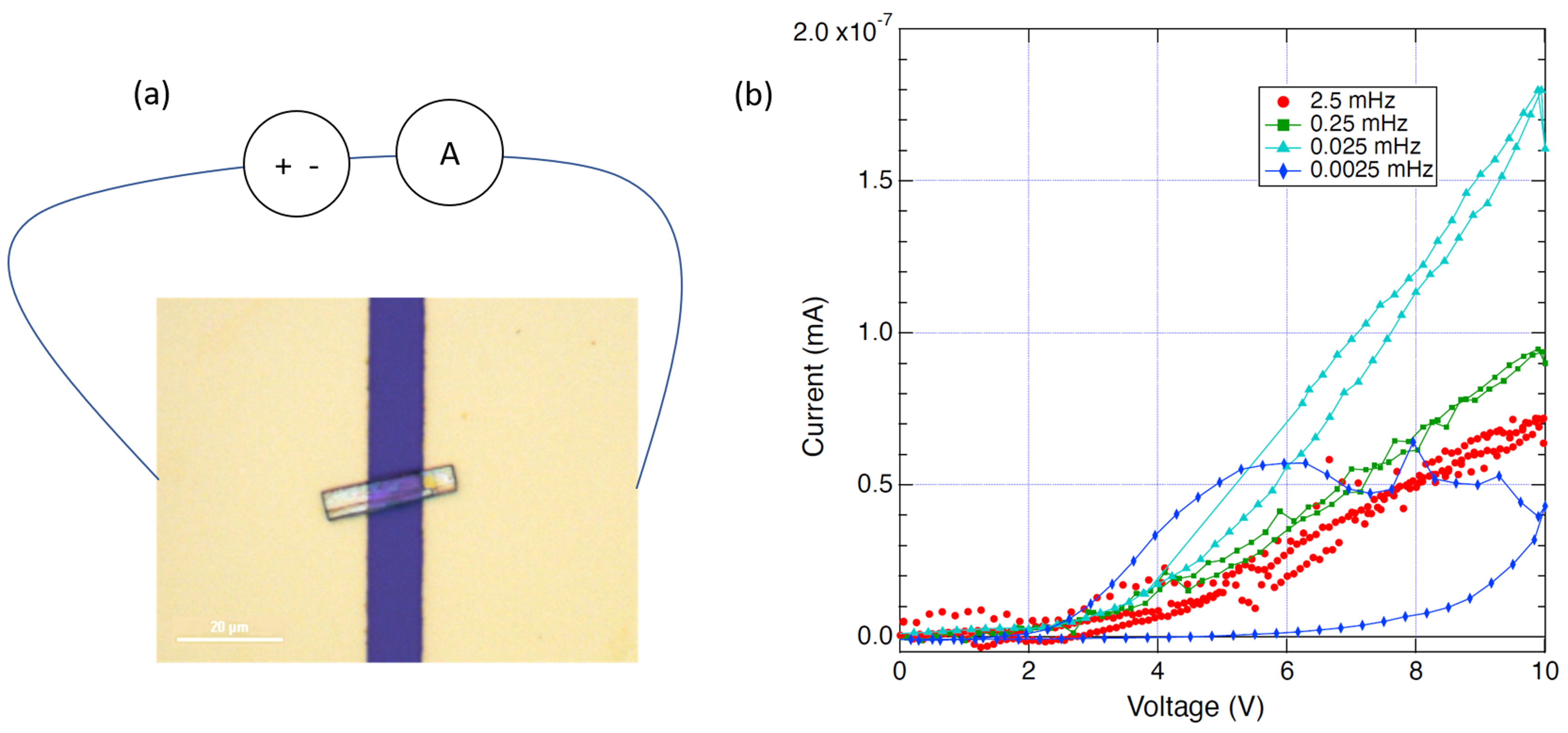}%
 \caption{\label{Dev}(a) Final device based on an exfoliated RTO nanocrystal (flake \#4) transferred on top of two patterned Au electrical contacts. The distance between the two electrodes is 10 $\mu m$. The purple layer corresponds to the insulating Si/SiOx substrate.  (b) I-V characteristics measured at room temperature for flake \#4 for various sweeping rates that are indicated by the corresponding frequencies. The behaviour is typical of a memristive device for which the I-V curves are single-valued above a certain frequency.}%
\end{figure*}

A single nanoflake was consequently tested for its electrical response, by fabricating a lateral two-terminal device and looking for an electrical response consistent with the bulk crystal response, in particular their memristive response as reported earlier\cite{PhysRevMaterials.1.032001}. 
The device is composed of two Ti/Au electrical connections at both edges of a single RTO nanocrystal. The fabrication process was entirely carried out in a controlled clean-room environment by standard micro- and nano-fabrication techniques. AZ5214E resist was used for the UV lithography as it is an image-reversal photoresist suitable for lift-off techniques. The developer was AZ326 MIF. The metallic contacts were achieved by evaporating thin Au films (50\,nm) on top of a thin Ti adhesion layer (1\,nm) onto a substrate of Si/SiO2 (300\,nm). The "hot pick-up" transfer method \cite{pizzocchero_hot_2016} was then used for the positioning of the exfoliated RTO nanocrystal in between the two metallic electrodes separated by a gap of 10 \,mum. An optical microscope image of a final device is depicted in Fig. \ref{Dev}a. 
Because of the reactivity of RTO with water, it was not possible to deposit the metallic contacts on top of an exfoliated RTO flake. It was necessary to transfer the nanocrystals on top of a pre-patterned substrate, although this design does not guarantee an optimal geometry of the contacts at the RTO/metal interface. 

Electrical measurements were performed at room temperature and ambient conditions using a SP300 potentiometer from Biologic. I-V characteristics are depicted in Fig.\ref{Dev}b for sample \#5  of dimensions 25 µm × 5 µm × 800 nm, using different voltage sweeping rates. These characteristics are typical of a memristive behavior \cite{chua_resistance_2011}: they appear pinched in zero, their area shrinks with increasing frequency and they are single-valued at high frequencies.
The maximum conductance value may be extracted from the left region of the blue curve or from the turquoise curve and corresponds to about $G= 2.5\times10^{-11}\,\mbox{S}$, while the minimal value is about 100 times smaller. 
Taking into account the dimensions of the sample, the maximal conductivity can be estimated to  $\sigma=5\times10^{-6}\,\mbox{S.cm}^{-1}$.  This value is in agreement with the expectations based on previous results  \cite{PhysRevMaterials.1.032001} although lower than the value obtained in the best activated bulk crystals that was about $10^{-4}\,\mbox{S.cm}^{-1}$ at room temperature. A conductivity increase of at least 4 orders of magnitude was observed in hydrated bulk RTO crystals with respect to dehydrated ones\cite{DESOUSACOUTINHO2021115630}.

There are two main differences compared to the results presented for bulk RTO crystals in Ref.\,\onlinecite{PhysRevMaterials.1.032001}. The first regards the intensity  of the electric field across RTO, since the range of applied voltage is the same, but in the present study the gap between the contacts is about 100 times smaller, meaning that the electric field is 100 times larger. Secondly, the contact geometry is also different. As a matter of fact, the nanocrystal was transferred on top of the pre-patterned metal contacts, but there was no direct deposition of the metal on top of RTO, so the electrical response is a combination of out-of-plane (along the c-axis) and in-plane (ab-plane) transport contributions while for RTO bulk crystal the contacts were aligned along opposite edges so that the contribution was mostly b-axis. While this difference in the geometry may explain the lower value of the conductivity, that is expected to be much weaker along the c-axis than along the ab-planes, the response carries some similarity, with a lobe developing at low frequency around 5\,V. The main difference is therefore the threshold frequency above which the I-V curve is single valued. For bulk RTO this transition was measured at around 50 mHz, whereas in the present study it is more around 0.01 mHz so about 5000 times smaller. This indicates that the memristive behaviour is extremely slowed down, which, could be presumably due to the extra-contribution of the c-axis.  
However despite these discrepancies, the RTO nanocrystals, at the scale investigated here, behaves qualitatively like a single RTO bulk crystal. 
These results therefore confirm that the RTO conduction properties are determined by the hydration level of the crystal and establish that memory effects due to ionic motion are preserved below micrometer scale.

The adhesive-tape technique for mechanical exfoliation\cite{doi:10.1073/pnas.0502848102} was investigated as a possible process to obtain thin $\mbox{Rb}_2\mbox{Ti}_2\mbox{O}_5$ flakes. Because of its lamellar structure as revealed in Fig. \ref{MEB}b, RTO represents a promising material for obtaining atomically thin layers. By using this technique, several crystals have been exfoliated and most of them present, at the end, a thickness close to 300\,nm. The limit of this technique has not yet been assessed and thinner flakes may be obtained.
Exfoliated RTO has been characterized through Raman spectroscopy measurements. A possible dependence of the collected Raman signal with respect to the RTO thickness has been studied. The results have shown that, the hydration level of the flakes, rather than the thickness, has a stronger influence on the spectra. 
Spectra of flakes indeed show similar features to those of "activated" crystals presented in Ref.\onlinecite{Federicci_The_2017}, pointing towards the role of water absorption\cite{DESOUSACOUTINHO2021115630} as the main cause of the changes with respect to as-grown crystals.

Finally, after device fabrication, electrical measurements have been conducted on one device connecting a single RTO nanocrystals. Interesting preliminary results have been obtained. In particular, conductivity values of about $5\,\mu \mbox{S.cm}^{-1}$, have been inferred.
While the overall memristive behavior is qualitatively similar, the slower RTO response to time-varying signals as well as the lower conductance presented in this work can be attributed to the different electrical geometry of the device. Indeed, the electrical contacts are lying below the RTO nanocrystal which gives an additional contribution from the c-axis with respect to the longitudinal geometry with conduction measured along the ab-plane, that was investigated in previous work on bulk RTO crystals. This must be further explored by depositing the metallic layer directly on top of the exfoliated RTO nano-crystals. 

What presented here was therefore a first approach to the investigation of $\mbox{Rb}_2\mbox{Ti}_2\mbox{O}_5$ at the sub-micrometer scale. A deep interest in carrying on with the analysis of its physical and electrical properties comes from the preliminary yet noticeable results obtained in this work, which have demonstrated in particular a preservation of the bulk memristive properties of this material at the nanoscale. Investigation of novel exfoliation techniques may open new possibilities in terms of shape, thickness, contacting geometry and local crystalline structure of RTO flakes or even to obtain nanosheets. Further research in this direction may even unveil new physical properties and give some more insights on the precise ions transport mechanism, still not fully understood.

\begin{acknowledgments}
C.B. and M.L.D.R. wish to acknowledge the support of  Stéphan Suffit, Pascal Filloux and Merdhad Rahimi during the micro and nanofabrication process. We also acknowledge Yann Gallais and Alexander Alekhin from the Raman-AFM platform (MPQ, Université Paris Cité) funded by ANR-18-IDEX-0001, IdEx Université Paris Cité. The work at LPEM was partly supported by PSL University through a maturation project in the framework of the "Investissements d'avenir" ANR program.

\end{acknowledgments}

\section*{Data Availability Statement}

The data that support the findings of this study are available from the corresponding author upon reasonable request.

\nocite{*}
\bibliography{valeriobib}

\end{document}